# $\mathcal{L}_1$ Adaptive Controller - Performance Analysis of the Inverse DC Gain Method

Dr. Sanchito Banerjee[1]

*University of Queensland, Brisbane, School of Mechanical and Mining Engineering, Queensland, 4072, Australia*

**This paper presents an analysis of the modified $\mathcal{L}_1$ adaptive control law. The performance of this control law is compared to the original control law. The modified $\mathcal{L}_1$ control law uses the DC gain of the transfer function of the closed loop plant dynamics. There is slight worsening of the controller performance. Furthermore, this analysis shows that provided that there is room for slight performance reduction, $\mathcal{L}_1$ adaptive control law can be used to control non-minimum phase systems without the use of a pole-zero cancellation technique. $\mathcal{L}_1$ adaptive control requires five assumptions before it can be applied. The stability of matched transmission zeros is no longer a condition that a system needs to meet as a result of this modification to the adaptive control law.**

## Nomenclature

| | |
|---|---|
| $A_M$ | = desired closed loop system matrix |
| $A_p$ | = plant model system matrix |
| $B_1$ | = input gain matrix |
| $B_2$ | = null space of $B_1$ |
| $c$ | = output matrix |
| $\dot{e}(t)$ | = tracking error dynamics |
| $e_P$ | = tracking error |
| $k_x$ | = Autopilot feedback gains |
| $L_P$ | = modification matrix in the state estimator |
| $m$ | = mass of vehicle, kg |
| $x$ | = state vector |
| $\hat{x}$ | = state estimate vector |
| $\hat{y}$ | = output estimate |
| $\gamma$ | = flight path angle, rad |
| $\zeta$ | = damping ratio |
| $\sigma$ | = input disturbances |
| $\hat{\sigma}$ | = estimate of error |

## I. Introduction

Over the past few decades, a considerable amount of funds have been directed to the research and development of hypersonic vehicles by the U.S. Air Force and NASA. Some of the main challenges of a hypersonic vehicle program from a control point of view are: (i) the large flight envelope,(ii) rapidly changing aerodynamic coefficients (iii) the uncertainties associated with them and (iv) the coupling between the structure and the aerodynamics of the vehicle [1-3]. A better understanding of these flight envelopes, rapidly changing aerodynamic coefficients and the associated uncertainties along a descent trajectory are the main motivations of this paper. Due to these complications, pole placement controller augmented with an $\mathcal{L}_1$ adaptive controller is utilised.

Scramjet powered rockets have the potential to simplify the launch of small satellites into orbit at a fraction of the cost of traditional rockets. The HyShot program outlined in [4-7] proved that a scramjet can produce thrust. This was on the descent trajectory of the ballistic trajectory. This means that the vehicle needs to be controlled during re-entry and post re-entry along a descent trajectory and deliver the scramjet to the right initial conditions. This study acts as a tool to thoroughly test the stability, robustness and performance characteristics of the controllers that could be used to deliver the scramjet to the right initial conditions. Moreover, this provides a certain baseline level of confidence when it is time for a test flight. Study and examples of the use of pole placement technique (used as the baseline controller in this study) in the case of hypersonic vehicle is placed in [8] (pp 404-407) and in [9-12].

---

[1] Adjunct Research Fellow, School of Mechanical and Mining Engineering, University of Queensland, Brisbane.

$\mathcal{L}_1$ adaptive control was initially developed to apply to flight control. The initial concept of $\mathcal{L}_1$ adaptive control was presented in [13-19]. The main aim of $\mathcal{L}_1$ adaptive control is to decouple the estimation and the control loops. This allows for fast adaptation without loss of robustness [20-26]. Literature depicting the application of $\mathcal{L}_1$ adaptive control are placed in [18, 20-22, 24, 25, 27-49] and are summarised here.

One of the practical applications of a full $\mathcal{L}_1$ adaptive controller is placed in [32], wherein a controller is designed for the NASA AirSTAR flight test vehicle. The $\mathcal{L}_1$ adaptive controller is applied to a subscale turbine powered Generic Transport Model (GTM). The controller presented directly compensates for matched as well as unmatched uncertainties. The main motive for this program was to test the performance of the aircraft and the pilots under adverse conditions. These conditions include unusual attitudes, surface failures and structural damage. One of the main outcomes of this program was to test the performance and the evaluation of the controller beyond the edge of the normal flight envelope. This is particularly important as beyond the pre-defined flight envelop the risk of vehicle loss is high due to limited knowledge of nonlinear aerodynamics beyond stall and the potential for high structural loads. The corresponding $\mathcal{L}_1$ controller is designed at a single operating point at the centre of the nominal flight envelope. The analysis is carried out for variable dynamics and damage cases in highly nonlinear areas of the flight envelope, in both piloted and non-piloted evaluations. As a result of these simulations, a theoretical extension of the $\mathcal{L}_1$ adaptive controller that accounts for the matched as well as unmatched dynamic uncertainties is presented. Two different sets of simulations are carried out: batch simulations and piloted simulations. The batch simulations are carried out on the nominal aircraft, the case where the entire rudder is missing, the case where left outboard trailing edge flap missing, loss of outboard left wing tip, loss of entire elevator from left stabilizer and loss of entire left stabilizer. For the case of piloted simulations: a static stability degradation, roll damping degradation, simultaneous static and roll degradation, high angle of attack capture task, sudden asymmetric thrust, and pilot induced upset and recovery cases are investigated. For all the aforementioned cases, the $\mathcal{L}_1$ adaptive controller performed well.

Griffin et. al in [20] presents a $\mathcal{L}_1$ adaptive control augmentation system to control the lateral/directional dynamics of X-29. This study presents a comparative study between the baseline Linear Quadratic Integral (LQI) controller and the augmented controller for the Single Input Single Output (SISO) and the Multi Input Multi Output (MIMO) setup. The augmentation system is once again utilised in the presence of unmatched uncertainties which may exhibit significant cross-coupling effects. An additional advancement is the addition of a high fidelity actuator model to the $\mathcal{L}_1$ architecture to reduce the uncertainties in the state predictor design. A predictable outcome of this study is the cancellation of the cross coupling effects when the multi-input multi output approach is used. It is also demonstrated that the $\mathcal{L}_1$ state predictor can be designed to match the nominal closed loop behaviour given a certain baseline controller. As expected under nominal conditions the adaptive element is minimal. However, in the presence of a failure, the augmentation system proves effective in regaining the nominal performance at the same time eliminating the undesired sideslip-to-roll coupling. The failure scenarios covered in this study are jammed rudder, jammed elevon and the reduction of the effect of cross coupling. A comparison of the SISO and the MIMO augmented systems in the presence of failures is placed in Fig. 1.

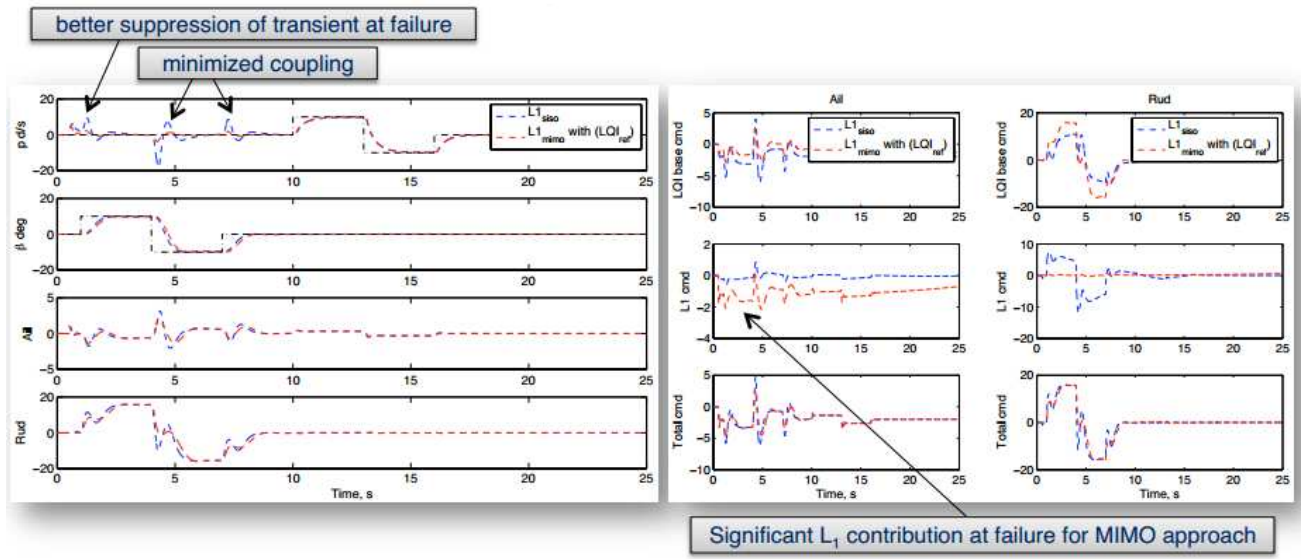

**Fig. 1 L1 Augmented Controller Performance Comparison [20]**

A $\mathcal{L}_1$ output feedback output controller architecture has been utilised to control the longitudinal dynamics of a missile and also for a flexible space Crew Launch Vehicle (CLV) in [22] and [35]. Kharisov et.al in [22] outlines the functionality and applicability of an $\mathcal{L}_1$ adaptive controller for a launch vehicle which operate often in very unforgiving and occasionally highly uncertain environments. The $\mathcal{L}_1$ output feedback controller architecture is utilised to control the low frequency structural modes. The paper also proposes steps to validate the adaptive controller performance. The theoretical performance bounds of $\mathcal{L}_1$ adaptive controller are verified through the nonlinear simulations. The results presented demonstrate that a single $\mathcal{L}_1$ adaptive output controller is able to handle statically unstable flexible plant with large parametric variations without addition of notch filters, without re-tuning for different flight conditions along the first stage trajectory and with guaranteed transient performance.

The output controller utilised to control the longitudinal dynamics of the missile autopilot is placed in [35]. The controller has satisfactory performance in the presence of parametric uncertainties and time-varying disturbances. A comparative study between the Linear Quadratic Regulator (LQR) and the Linear Quadratic Gaussian (LQG) with Loop Transfer Recovery (LTR) design with the $\mathcal{L}_1$ adaptive controller is presented. And the simulations results present the clear benefits of using the $\mathcal{L}_1$ controller. A further study of the application of an $\mathcal{L}_1$ adaptive controller to a missile in placed in [28]. A $\mathcal{L}_1$ control augmentation is designed based on a novel autopilot structure for a highly, agile, tail-controlled missile. The autopilot is designed to control the pitch and paw accelerations while maintaining a desired roll angle. The $\mathcal{L}_1$ adaptive augmentation is implemented and utilised to compensate for the undesirable effects of modelling uncertainties. This augmentation is successful in increasing the robustness of the baseline controller in the presence of significant plant uncertainties. This paper is successful in showing that, despite the reduced complexity, the $\mathcal{L}_1$ augmented controller is able to increase the robust performance when compared to an Model Reference Adaptive Control (MRAC) augmentation placed in [50].

Leman in [24] presents the $\mathcal{L}_1$ augmented controller for the longitudinal and the lateral/directional dynamics for a X-48B aircraft. The piecewise constant adaptive law is utilised to estimate the matched and unmatched uncertainties. SISO channels are utilised on all the channels: longitudinal, lateral and directional. The performance of the baseline controller is compared to the augmented controller in the presence of control surface failures. An additional conclusion to be drawn from this study is the lack of high frequency components in the control surface deflections. The $\mathcal{L}_1$ augmented controller improves the performance of the baseline controller under nominal conditions across the entire envelope. In the event of a failure, the $\mathcal{L}_1$ adaptive controller adapts to recover desired aircraft performance and provide a predictable response to pilot inputs.

A full $\mathcal{L}_1$ adaptive controller is presented by Prime in [51] to control the longitudinal dynamics of a waverider class hypersonic vehicle where the aerodynamics are based on NASA's Generic Hypersonic Aerodynamics Model Example (GHAME). To test the robustness and the performance properties of the $\mathcal{L}_1$ adaptive controller, pull-up trajectory simulations are performed using several different simulation scenarios with no change to the controller configuration. The simulation scenarios covered in the paper are: (i) Nominal conditions, (ii) Ramp reduction from nominal elevator effectiveness, $M_{\delta e}$ to $0.5 M_{\delta e}$ starting at an altitude of 28km over 5 seconds, (iii) Step reduction of the elevator effectiveness at an altitude of 28km, (iv) Reduced static margin, (v) Controller processing delay of 10ms, and (vi) Controller processing delay of 20ms. Projection based adaptive laws are used to compensate for the matched and the unmatched uncertainties. The adaptive controller is shown to be an effective control mythology for a re-entry vehicle. However, for some of the off-nominal conditions there was reduced performance.

## II. $\mathcal{L}_1$ Controller Design and Validation

The philosophy behind this $\mathcal{L}_1$ adaptive controller is to obtain a state feedback controller, estimate the uncertainties in the system and compensate for these uncertainties within the bandwidth of a predesigned low pass filter. The filter ensures that the controller remains in the low frequency range in the presence of fast adaptation and large reference inputs [23]. This adaptive control scheme is chosen for the following reasons [20-22, 24, 28, 32, 33, 52, 53]:

I. Decoupling of the rate of adaption and robustness – this is achieved with the help of the low pass filter.
II. Guaranteed, bounded away from zero time delay margin.
III. Guaranteed transient performance for a system's input and output signal, without high gain feedback or enforcing persistent excitation type assumptions.
IV. Reduced cost for the validation and verification process: this is achieved with the help of the analysis that is carried out on the $\mathcal{L}_1$ reference system and the $\mathcal{L}_1$ design system.

A model of the augmented controller is placed in Fig. 2. The closed loop plant includes the baseline controller.

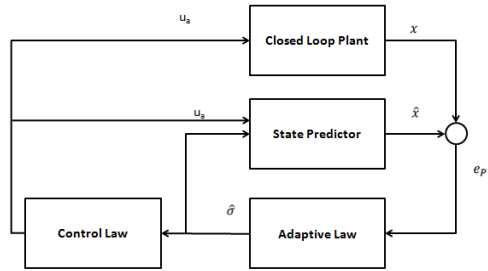

**Fig. 2 $\mathcal{L}_1$ Controller Model**

In order to realise control objectives, some further assumptions about the uncontrolled plant dynamics are made, while applying the $\mathcal{L}_1$ piecewise constant adaptation method. These include [10, 12, 23]

ASSMP 1  **Boundedness of $f_i(t, 0)$:** $\exists B_i > 0$, such that $\|f_i(x_P, x_I, t)\|_\infty < B_i, i = 1,2$ holds for all $t \geq 0$. ($f_i(t, 0)$ only assumes limited values → typically fulfilled for stable mechanical systems; $B_i$ is an upper bound for the uncertainties – matched and unmatched).

ASSMP 2  **Semiglobal Lipschitz Condition:** For arbitrary $\delta > 0, \exists K_{1\delta}, K_{2\delta}$, such that: $\|f_i(X_1, t) - f_i(X_2, t)\|_\infty \leq K_{i\delta}\|X_1(t) - X_2(t)\|_\infty, i = 1,2$ for all $\|X_j(t)\|_\infty \leq \delta, j = 1,2$, uniformly in $t$.

ASSMP 3  **Stability of Unmodelled Dynamics:** The $x_I$ - dynamics are BIBO stable both with respect to initial conditions $x_{I0}$ and the input $x_P(t)$, i.e.: $\exists L_Z, B_Z > 0$, such that for all $t \geq 0$, $\|y_{I_t}\|_{L_\infty} \leq L_Z \|x_P\|_{L_\infty} + B_Z$

**ASSMP 4  Partial Knowledge of the System Input Gain:** (i) $\Lambda \in \mathbb{R}^{m \times m}$ is assumed to be unknown (non-singular) strictly row-diagonally dominant matrix with $sgn(\Lambda_{ii})$ known, i.e. $|\Lambda_{ii}| > \sum_{j \neq i}|\Lambda_{ij}|$. (ii) Also assume that there exists a known compact convex set $\Omega$, such that $\Lambda \in \Omega \in \mathbb{R}^{m \times m}$, and the nominal system input gain $\Lambda_0 \in \Omega$ is known. By knowing $B_1$ we know the rough direction in the state space which are produced by the inputs. With the help of $\Omega$ the effectiveness of the inputs acting in the direction of $B_1$ are defined. This effectiveness is unknown, but assumed to be within a known interval given by $\Omega$ which also contains the nominal effectiveness $\Lambda_0$.

**ASSMP 5  Stability of Matched Transmission Zeroes:** The transmission zeroes of the transfer matrix $H_{trans} \triangleq C_P(sI_n - A_M)^{-1}B_1$ lie in the open left half plane (the system is minimum phase → the initial response to inputs has the same direction as the steady state response).

The control objective is to design an augmentation law that compensates for the matched and unmatched uncertainties on the output of the system $y$, and ensures that the output tracks the desired output $y_M$, given a certain reference signal $r$. This is to occur both in transient and steady state while all signals remain bounded.

**Vehicle Plant Dynamics:**

The plant dynamics with the baseline controller of a system is.

$$\dot{x}(t) = A_M(t)x(t) + B_1(t)K_r r(t) + B_1(t)\big(u_a(t) + f_1(x_P, x_I, t)\big) + B_2(t)f_2(x_P, x_I, t)$$
$$x(0) = x_0, \quad y = c^T x(t) \tag{1}$$

where $A_M$ is the closed loop plant model which includes the baseline controller. $B_1$ is the control input matrix. $B_2$ is chosen such that $B_1^T B_2 = 0$ and $Rank[B_1 B_2] = n$. $B_1$ and $B_2$ are the control gain matrices for the matched and the unmatched uncertainties. $f_1$ includes the matched uncertainties and $f_2$ includes the unmatched uncertainties. Matched uncertainties in a system are represented in the following form: $\dot{x} = A_M x(t) + B_1\big(u(t) + f(x,t)\big)$, where $f(x,t)$ is an unknown nonlinear function that is within the span of the control input $u(t)$. Uncertainties in the state space matrices that cannot be represented in this form are the unmatched uncertainties. $K_r \triangleq -\frac{1}{c^T A_M^{-1}(t) B_1(t)}$ is the feedforward gain. $u_a(t)$ is the adaptive control signal.

**Desired Plant Dynamics:**

The desired closed loop plant dynamics is

$$\dot{x}(t) = A_M(t)x(t) + B_1(t)K_r r(t), \; x(0) = x_0$$
$$y_M = c^T x(t) \tag{2}$$

**State Predictor:**

$$\dot{\hat{x}}(t) = A_M(t)\hat{x}(t) + B_1(t)K_r r(t) + B_1(t)\big(u_a(t) + \hat{\sigma}_1(t)\big) + B_2(t)\hat{\sigma}_2(t) + L_P(t)\tilde{x}(t), \; \hat{x}(0) = x_0$$
$$\hat{y} = c^T \hat{x}(t) \tag{3}$$

where $\hat{\sigma}_1(t)$ and $\hat{\sigma}_2(t)$ are the estimates of the matched and the unmatched uncertainties respectively. The estimates of these uncertainties are provided by the $\mathcal{L}_1$ piecewise constant update law. $L_P$ is a matrix used to assign faster poles for the prediction error dynamics and $\dot{\tilde{x}}$ is $\dot{\hat{x}} - \dot{x}$ and $\tilde{x} = \hat{x} - x$ is the predictor error.

**Control Law:**

The modified adaptive signal, as used in [12], is (written in matrix form due to the different bandwidths of the low pass filter):

$$u_a(s) = [-C_2(s) \quad -C_1(s)] \begin{bmatrix} \frac{1}{-(c^T(A_M)^{-1}B_1)} H_2(s)\hat{\sigma}_2(s) \\ \hat{\sigma}_1 \end{bmatrix} \quad (4)$$

The main advantage of this implementation is that the inversion of the system dynamics is not needed. This means that this form of the equation can be implemented for a nonminimum phase LTV system. The original adaptive control law is as follows:

$$u_a = C(s)\left(-H_1^{-1}H_2\hat{\sigma}_2(s) - \hat{\sigma}_1(s)\right) \quad (17)$$

However, for a nonminimum phase system $H_1^{-1}(s)$ is unstable. Scheduling $H_1^{-1}(s)H_2(s)$ can lead to the switching itself being unstable. Therefore, it needs to be carried out carefully. However, the scheduling the pole-zero cancellation (to cancel the unstable pole) can lead to unpredictable closed loop dynamics. Therefore, the inverse of the dc gain is utilised to carrying out the LTV implementation of the transfer function. This method was first proposed by Che et. al. in [54].

**Low Pass Filter C(s) :**

The structure of the first order low pass filter is given by

$$C_i(s) \triangleq \frac{\omega_b}{s + \omega_b}; \text{where } i = 1,2 \quad (18)$$

where $\omega_b$ is the bandwidth of the low pass filter. In the case of the hypersonic glider, two first order filters of different bandwidths are implemented for the matched and the unmatched uncertainties. The bandwidth of the low pass filters is a direct result in the nature of uncertainties that make up the matched and unmatched uncertainties.

**Piecewise Constant Update Law:**

The piecewise constant update law that calculates the estimated values of the uncertainties is

$$\hat{\sigma}(iT_S) = -B^{-1}\Phi^{-1}(T_S)\mu(T_S), t \in [iT_S, iT_S + T_S] \quad (19)$$

where:
$$\hat{\sigma}(iT_S) = \begin{bmatrix} \hat{\sigma}_1(iT_S) \\ \hat{\sigma}_2(iT_S) \end{bmatrix} = \begin{bmatrix} \text{matched uncertainties} \\ \text{unmatched uncertainties} \end{bmatrix}$$
$$\Phi(T_S) \triangleq A_S^{-1}(e^{A_S T_S} - I_n), \quad \text{where } A_s = L_P + A_M$$
$$\mu(T_S) = e^{A_S T_S}\tilde{x}_P(iT_S)$$
$$\tilde{x}_P = \hat{x}_P - x_p$$
$$B = [B_1 B_2] \text{ where } B_2 \text{ is the null space of } B_1 \text{ and } B \text{ is full rank}$$
$$T_S = \text{sampling period of the model}$$

The piecewise constant update law can be rewritten from an implementation point of view. Although the piecewise constant law has been developed (from a derivation point of view) for a LTI system, in a discretized system, every time step the $A_S$ is taken as a constant. Therefore, the existing piecewise constant law can still be used for a LTV augmentation setup.

$$\hat{\sigma}(iT_S) = (-B^{-1}\Phi^{-1}(T_S)e^{A_S T_S})\tilde{x}(iT_S) \quad (20)$$

## III. Simulation, Results and Analysis - Longitudinal Dynamics

This section of the controller design looks at the validation of the modified $\mathcal{L}_1$ augmented control law as applied to a simplified case. This test is carried out to test the performance of the controller for a simple SISO LTI test case before being applied to the hypersonic vehicle. This is an important step in the design process as it gives confidence in the fact that the controller does not just work for an isolated case. Rather, this section acts as a way of proving that the controller can be applied to various classes of linear systems. There are three different avenues to validate a controller. They are:

- Flight Test
- Hardware-In-the-Loop Simulations
- Simulation based Validation

Due to the limitation of the project in terms of time and the financial situation, a simulation based validation methodology is used. Results in this section gives the confidence before the modified control law is applied to the hypersonic test case. The SISO system considered as a test case is placed here.

$$A_m = \begin{bmatrix} -10 & -50 \\ 1 & 0 \end{bmatrix}$$
$$B_m = \begin{bmatrix} 2000 \\ 0 \end{bmatrix}, \quad C = \begin{bmatrix} 0 & 1 \end{bmatrix}$$

The system considered for this validation test is of the following form

$$\dot{x}(t) = A_m x(t) + B_1 K_g r(t) + B_1 u_a(t) + B_1 f_1(x_p, x_I, t) + B_2 f_2(x_P, x_I, t) \tag{21}$$
$$y(t) = C x(t), \quad x(0) = x_0$$

The feedforward gain $K_g = -(CA_m^{-1}B_m)^{-1}$. For this case $K_g$ is 0.025. $r(t)$ is a step command and a ramp command with a gradient of one. The uncertainties added to the system are static uncertainties: $f_1 = 0.05$ and $f_2 = 0.001$. The performance of the system without the augmentation, with the original implementation of the augmentation (placed in Eq. 17) and the modified augmented control law (placed in Eq. 16).

### A. Nominal Case - Step and Ramp Input

The case presented here outlines the performance without the augmentation for the step and the ramp command. The matched and the unmatched uncertainties are set to zero. Both the step and the ramp command is introduced into the system at $T = 2$. For both the cases, the system tracks the input without any steady state error. for the case of the ramp input, there is a slight delay in the tracking variable. This is consistent with systems tracking time varying commands.

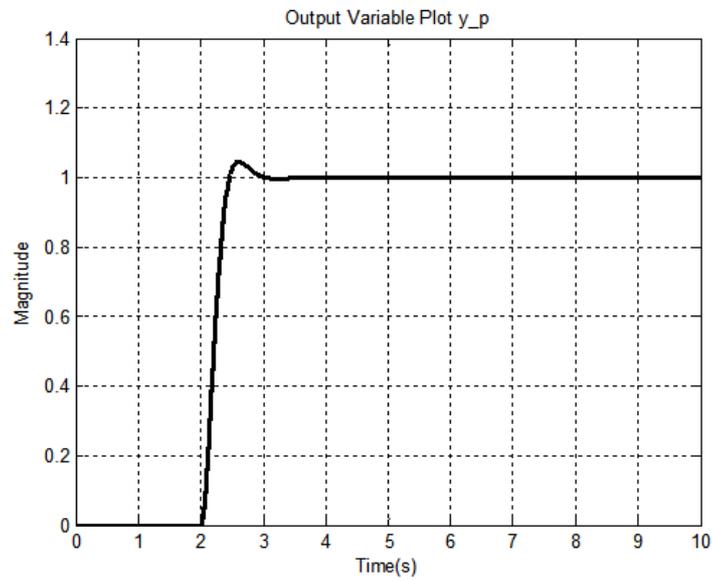

**Fig. 3 System Output - Step Response**

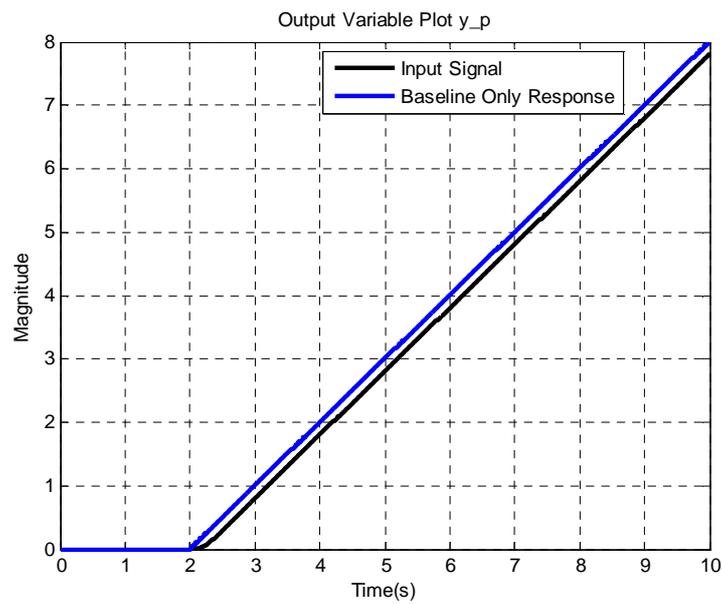

**Fig. 4 System Output - Ramp Input**

### B. Uncertainties - Step Input

The performance of the system in the presence of matched and unmatched uncertainties is presented. The uncertainties are set to $f_1 = 0.05$ and $f_2 = 0.001$. The behaviour of the tracking variable is placed in Fig. 5. Although stable, the performance is severely degraded due to the presence of uncertainties for the case when the augmentation is turned off.

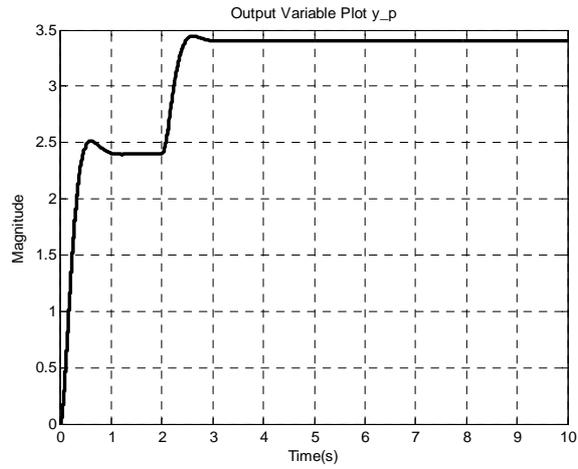

**Fig. 5 Response to Step Input with Uncertainties**

In order to also determine the transient behaviour of the modified augmented control law as compared to the original, only the matched component of the augmentation law is utilised to test the performance and compare it to the complete augmentation setup. Therefore, for the next test case the adaptive control law is $-C(s)\hat{\sigma}(s)$. The adaptive signal is placed magnitude placed in Fig. 7 depicts the control signal that is compensating for the uncertainties in the system. However, it is seen in Fig. 6, that the performance of the system is not restored. This is due to the unmatched uncertainties being present in the system, and the unmatched component of the adaptive control law being turned off.

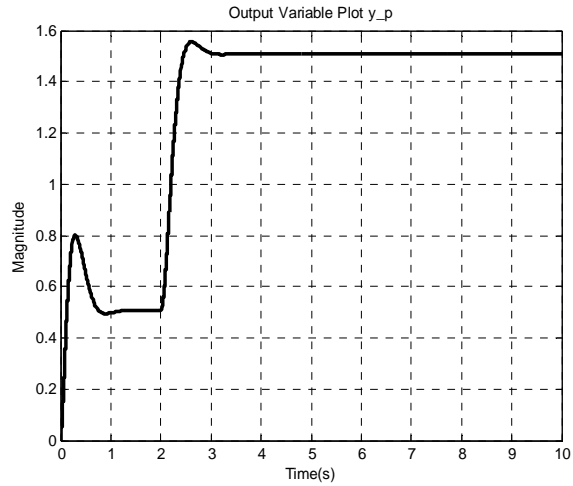

**Fig. 6 Performance of System - Matched Component of Adaptive Law**

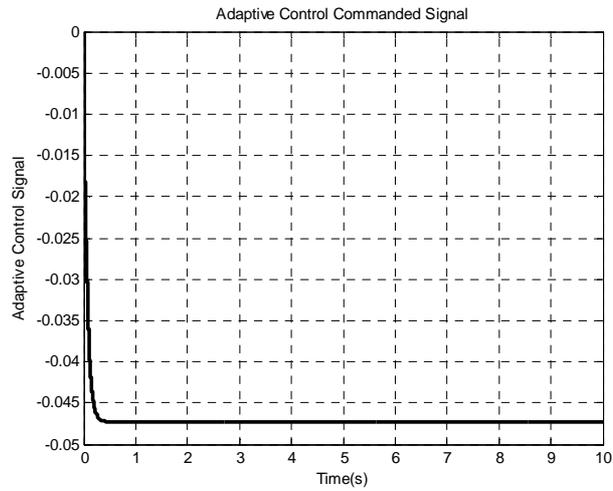

**Fig. 7 Adaptive Signal Magnitude**

Due to the fact that the unmatched contribution of the control law is not considered in the above diagram, there is a steady state error that is introduced into the tracking variable. This test adds as a further test as to whether the modified control law effects the performance of the tracking variable during the transient period. The plots of the original control law (which involves inverting the system dynamics), and the modified control law (which involves using the inverse DC gain of the system dynamics) are placed in Fig. 8 and Fig. 9. The modified control law exhibits a slightly higher overshoot and a slightly slower response as compared to the original adaptive control law. However, the advantage of avoiding the inversion of the system dynamics is seen as enough of an advantage to allow this slight worsening in the performance.

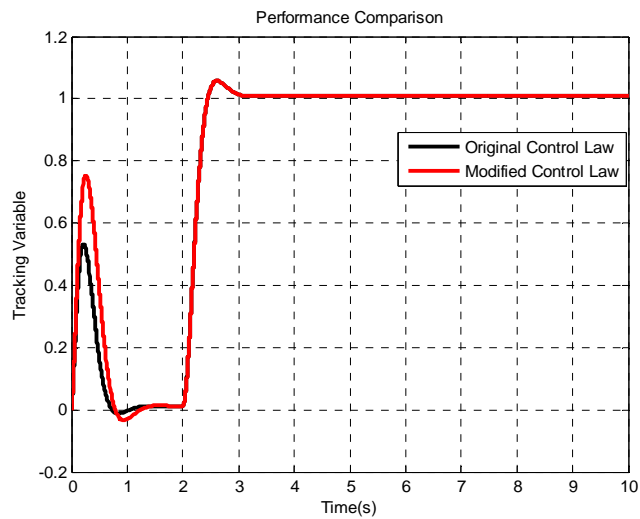

**Fig. 8 System Performance - Unmatched Component Turned On**

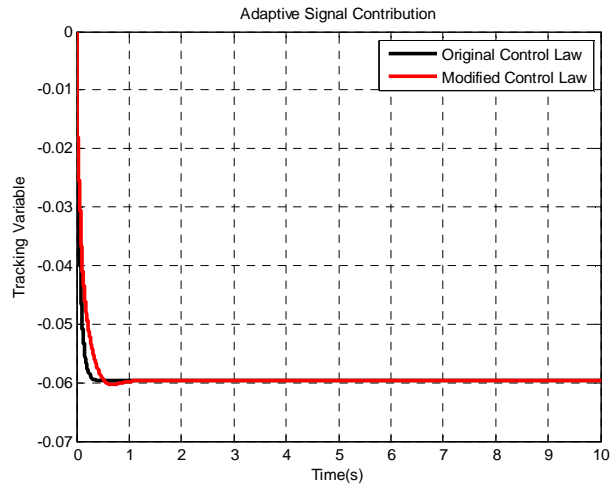

**Fig. 9 Adaptive Signal Contribution (Original and Modified)**

The steady state behaviour of both the adaptive control laws is identical. However, the transient behaviour shows a difference. The modified controller with its inverted DC gain of the system dynamics shows a slower response. This can be seen by the behaviour of the adaptive signal. The red line in Fig. 9 reaches it steady state value a bit later than the black line. Furthermore, the overshoot in Fig. 8 for the modified adaptive control is more than the original adaptive control law.

### C. Uncertainties - Ramp Input

In the presence of uncertainties, the performance of the system to a ramp input is placed in Fig. 10. The black line in the figure is the commanded signal. Once again, although the response is stable, the performance is significantly degraded due to the presence of uncertainties without the $\mathcal{L}_1$ augmentation.

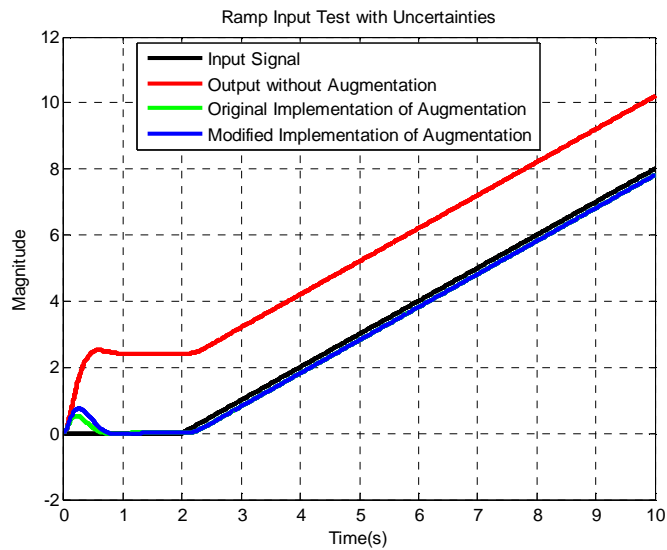

**Fig. 10 System Performance**

The blue and the green lines show the performance of the system with the original and the modified adaptive control law. Once again the steady state performance for both the augmentation is identical. The transient behaviour is once again similar as to the once seen with the step input. The transient behaviour is once again only slightly worse at the very start of the simulation.

## IV. Conclusion

This paper presented an analysis of the modified $\mathcal{L}_1$ adaptive control law. The performance of this control law is compared to the original control law. The modified $\mathcal{L}_1$ control law uses the DC gain of the transfer function of the closed loop plant dynamics. There is slight worsening of the controller performance. Furthermore, this analysis shows that provided that there is room for slight performance reduction, $\mathcal{L}_1$ adaptive control law can be used to control non-minimum phase systems without the use of a pole-zero cancellation. $\mathcal{L}_1$ adaptive control requires five assumptions before it can be applied. The stability of matched transmission zeros is no longer a condition that a system needs to meet as a result of this modification to the adaptive control law.